\documentstyle[12pt]{article}
\begin{document}
\baselineskip=12pt
\pagestyle{empty}
\today \\
\bigskip
\bigskip
\baselineskip=24pt
\begin{center}
\begin{large}
{\bf GROWTH LAWS FOR PHASE ORDERING}\\
\end{large}
\bigskip
\medskip
{\em A. J. Bray and A. D. Rutenberg} \\
\medskip
Department of Theoretical Physics \\
The University, Manchester M13 9PL \\
bray@mphv2.phy.man.ac.uk, ar@mphv2.phy.man.ac.uk \\
\bigskip
\medskip
{\bf ABSTRACT }
\end{center}
We determine the characteristic length scale, $L(t)$, in
phase ordering kinetics for both scalar and vector fields,
with either short- or long-range interactions, and with or
without conservation laws. We obtain $L(t)$ consistently by
comparing the global rate of energy change to the energy dissipation
from the local evolution of the order parameter. We derive
growth laws for  $O(n)$ models, and our results can be
applied to other systems with similar defect structures.

\bigskip
cond-mat/9303011

\newpage
\pagestyle{plain}
\pagenumbering{arabic}

Systems quenched from a disordered phase into an ordered phase
do not order instantaneously. Instead, the length scale of
ordered regions grows at a characteristic rate as different broken symmetry
phases compete to select the ground state. During this process,
energy is dissipated and topological defects, if present, are destroyed.
Traditionally, systems with scalar order parameters, such as
binary alloys and Ising models, have been studied \cite{Gunton83,Mason93}.
In such systems domains of both phases grow and the intervening
domain walls, which are the characteristic topological defects of
scalar systems, shrink in total area; the energy of the system is dissipated
as the total area of domain wall  decreases.  Recently, there has been
a growing interest in systems with vector and more complex
order parameters which have defect structures such as
lines, points, and textures. An exciting series of
experiments \cite{Wong92,Pargellis92} and
simulations \cite{Toyoki91,Blundell92,Mondello93,Siegert93,Yurke93} has
been performed to explore these systems. Of particular interest is the
time dependence, $L(t)$, of the characteristic
length scale. This growth law provides a basis for further analysis and
a fundamental test of approximation schemes. Growth laws have been obtained
in individual cases \cite{Lifshitz61,Lifshitz62,Bray89,Newman90}, but the
variety of possible systems has prevented the development of a unified
approach.  Renormalization group (RG) ideas \cite{Bray89,Bray93b} have been
used to predict the leading power-law dependence of growth laws,
but are not easy to apply in systems without conservation laws.
An approach to calculating growth laws, including logarithmic factors,
 for a wide variety of systems would be welcome.

In this Letter we  determine the characteristic length scale, $L(t)$, for
quenched systems with both scalar and vector fields,
with either short- or long-range interactions, and with or without conservation
laws. We base our approach on the energy dissipation that occurs as the
system relaxes towards its ground state.
We evaluate the energy dissipation by considering topological defects,
which dominate the dynamics when they exist.
We compare the global rate of energy change to the energy dissipation from the
local evolution of the order parameter. For systems with a single
characteristic scale describing correlations as well as dissipation
we self-consistently determine the growth law, $L(t)$.

Our results are summarized in tables 1 and 2. We find that conservation
laws ($\mu$) and long-range interactions ($\sigma$)
are irrelevant to the growth law below and above,
respectively, values determined by the number of components ($n$) in the
order parameter of the system. At the marginal values, logarithmic factors
are introduced. The growth laws obtained are independent of the
spatial dimension of the system. Our results should be compared to growth laws
obtained by other means. Heuristic arguments yield growth
laws of $L \sim t^{1/2}$  for
non-conserved \cite{Lifshitz62} and $L \sim t^{1/3}$ for conserved
\cite{Lifshitz61} scalar order parameters. RG arguments \cite{Bray89}
reproduce the $ t^{1/3}$ law for conserved scalar fields,  predict
$L \sim t^{1/4}$ for a conserved vector field, and also treat conserved
systems with long-range interactions \cite{Bray93b}.
Simulations \cite{Toyoki91} suggest a $L \sim t^{1/2}$ growth law
for non-conserved vector fields, and simulation \cite{Blundell92} and
experiments \cite{Wong92} obtain the same law for
the related tensor fields that describe nematic
liquid crystals.  This result has also been obtained by approximate
 treatments for general $n$ \cite{Liu92} and verified in the large-$n$ limit
\cite{Newman90}. Our approach provides a unified theoretical basis
for the leading power-law behavior suggested by these results,
and also determines any logarithmic factors.

The dynamic scaling hypothesis describes most systems
in the late stages of growth \cite{Furukawa85}.
Accordingly, the correlation function of the order parameter, $C({\bf r},t) =
\left< \vec{\phi}({\bf x},t)\cdot \vec{\phi}({\bf x}+{\bf r},t) \right>$,
with an average over initial conditions,
should exhibit the scaling form $C({\bf r},t) = f(r/L(t))$,
with a single characteristic length scale $L(t)$.
Fourier transforming this, we obtain the scaling form for the
structure factor $S({\bf k},t) = \left< \vec{\phi}_{\bf k}(t) \cdot
\vec{\phi}_{\bf -k}(t)\right>$,
\begin{equation}
\label{EQN:STRUCT}
S({\bf k},t) = L(t)^d\,g(kL(t))\ ,
\end{equation}
where $d$ is the spatial dimension.
A generic energy functional of such a system, with
an $n$-component field, $\vec{\phi}(\vec{x})$, is
\begin{equation}
\label{EQN:HAMILTONIAN}
H[\vec{\phi}] = \int d^d x\,[ (\nabla \vec{\phi})^2 + V(\vec{\phi}) ]\ ,
\end{equation}
where $V(\vec{\phi})$ is a ``mexican hat'' shaped potential such as
$V(\vec{\phi})= ( \vec{\phi}^2 -1)^2$. The
equation of motion for the ordering kinetics of
the Fourier components $\vec{\phi}_{\bf k}$ \cite{Hohenberg77} is
\begin{equation}
\label{EQN:DYNAMICS}
\partial_t \vec{\phi}_{\bf k} =
-k^{\mu}\,(\partial H/\partial \vec{\phi}_{\bf -k}),
\end{equation}
where we work at temperature $T=0$, with no thermal noise,
since $T$ appears to be  an `irrelevant variable' for ordering
kinetics within the ordered phase \cite{Bray89,Newman90}.
The conventional non-conserved (model A dynamics) and conserved
(model B dynamics) cases are $\mu=0$  and $\mu=2$, respectively.

Integrating the rate of energy dissipation from each Fourier mode, and
then using the equation of motion (\ref{EQN:DYNAMICS}), we find
\begin{eqnarray}
\label{EQN:LHSRHS}
d\epsilon/dt &=& \int_{\bf k} \left< (\partial H/\partial\vec{\phi}_{\bf k})
        \cdot \partial_t \vec{\phi}_{\bf k} \right>  \nonumber \\
&=& - \int_{\bf k} k^{-\mu}\,
        \left< \partial_t \vec{\phi}_{\bf k} \cdot
        \partial_t \vec{\phi}_{\bf -k}
        \right>\ ,
\end{eqnarray}
where $\epsilon = \left< H \right>/V$ is the mean energy density, and
$\int_{\bf k}$ is the momentum integral $\int d^d k /(2\pi)^d$. We will relate
the scaling  behavior of both sides of (\ref{EQN:LHSRHS}) to
that of appropriate integrals over the structure factor, $S({\bf k},t)$.
Either the integrals converge and the dependence on the scale $L(t)$ can
be extracted using the scaling form (\ref{EQN:STRUCT}), or a divergent
contribution dominates the integral. Any divergence will
occur because of small-scale structure at $kL \gg 1$, since no
structure has formed yet for $kL \ll 1$. This small-scale
structure will only come from topological defects, so we expect
$S({\bf k},t)$ to be proportional to the density of defect core,
$\rho_{\rm def} \sim L^{d-n}/L^d \sim L^{-n}$ for $kL \gg 1$ \cite{Bray93a}.
The scaling form (\ref{EQN:STRUCT}) then implies
\begin{equation}
\label{EQN:POROD}
S({\bf k},t) \sim L^{-n}\,k^{-(d+n)}\ ,\ \ \ \  kL \gg 1,
\end{equation}
which is a generalized Porod's law for $n \leq d$ \cite{Bray91,Toyoki92}.

We first calculate the scaling behavior of the energy density, $\epsilon$,
which  is captured by that of the gradient term in (\ref{EQN:HAMILTONIAN}):
\begin{eqnarray}
\label{EQN:GRADSQUARE}
\epsilon &\sim& \left< (\nabla \vec{\phi})^2 \right> \nonumber \\
         &=& \int_{\bf k} k^2\,L^d\,g(kL)\ ,
\end{eqnarray}
where we have used the scaling form (\ref{EQN:STRUCT})
for the structure factor.   If the integral converges, then a simple
change of variables yields $\epsilon \sim L^{-2}$.
If the integral diverges for $kL \gg 1$, then we
use Porod's law (\ref{EQN:POROD}) and impose
a cutoff at $k \sim 1/\xi$, where $\xi$ is the defect core size.
We obtain \cite{Toyoki92}
\begin{eqnarray}
\label{EQN:E}
\epsilon & \sim & L^{-n}\,\xi^{n-2}\ ,\ \ \ \ \ \ \ \ \ \ \ \ n<2\ ,\nonumber
\\
         & \sim & L^{-2}\,\ln(L/\xi)\ ,\  \ \ \ \ \ \ \ n=2\ ,\nonumber \\
         & \sim & L^{-2}\ ,\ \ \ \ \ \ \ \ \ \ \ \ \ \ \ \ \ \ n>2\ .
\end{eqnarray}
We see that the energy is dominated by the defect core density $\rho_{\rm def}$
for $n<2$, by the defect field at all length scales for $n=2$,
and by variations of the order parameter at scale $L(t)$ for $n>2$.

We now evaluate the right side of (\ref{EQN:LHSRHS}) in a similar way.
Using the scaling assumption for the two-time function,
$\left< \vec{\phi}_{\bf k} (t) \cdot \vec{\phi}_{-{\bf k}} (t') \right>
= k^{-d} g(kL(t),kL(t')\,)$, we find
\begin{eqnarray}
\label{EQN:TWOTIMES}
        \left< \partial_t \vec{\phi}_{\bf k} \cdot \partial_t
        \vec{\phi}_{-{\bf k}} \right> &=& \left. \frac{\partial^2}
        {\partial t \partial t'} \right|_{t=t'}
        \left< \vec{\phi}_{\bf k}(t) \cdot \vec{\phi}_{-{\bf k}}(t') \right>
        \nonumber \\
        &=& (dL/dt)^2 L^{d-2} h(kL).
\end{eqnarray}
When the momentum integral on the right of (\ref{EQN:LHSRHS}) converges,
we obtain, using (\ref{EQN:TWOTIMES}),
$d\epsilon/dt \sim - (dL/dt)^2 L^{\mu-2}$. However, if the integral
diverges we need to know the behaviour of (\ref{EQN:TWOTIMES}) for $kL \gg 1$.

In this regime, where the important structure is
short compared to the typical interdefect spacing, the scaling
functions are determined by the field of a single defect \cite{Bray93a},
and the defect field will be comoving with the defect core,
\begin{equation}
\label{EQN:COMOVING}
\partial_t\vec{\phi} = \vec{\omega}\times\vec{\phi} -
{\bf v} \cdot \nabla\vec{\phi}\ . \nonumber
\end{equation}
Here, ${\bf v}$ is the velocity of the defect and $\vec{\omega}$
represents rotations for vector fields. For $kL \gg 1$, the
velocity term dominates because the rotation scales as
$\omega \sim (dL/dt)/L$ while the gradient term involves the short-scale
structure, ${\bf v} \cdot \nabla \sim (dL/dt)\, k $.
In estimating $v \sim (dL/dt)$, we are assuming that dissipation occurs
by the evolution of defects of scale $L(t)$.
Using $\partial_t \vec{\phi}_{\bf k} \sim (dL/dt) \, k \, \vec{\phi}_{\bf k}$
and Porod's law (\ref{EQN:POROD}), we have
\begin{equation}
\label{EQN:RHSLARGEK}
\left< \partial_t \vec{\phi}_{\bf k} \cdot
\partial_t \vec{\phi}_{\bf -k}\right>
 \sim  L^{-n}\,k^{-(d+n-2)}\,(dL/dt)^2\ ,\ \ \ \ \ \ kL \gg 1\ .
\end{equation}
Using (\ref{EQN:RHSLARGEK}) in the right of (\ref{EQN:LHSRHS})
shows that  the integral is convergent for $kL \gg 1$ when $n+\mu>2$.
Otherwise the integral is dominated by $k$ near the upper cut-off $1/\xi$.
We obtain
\begin{eqnarray}
\label{EQN:RHS}
\int_{\bf k} k^{-\mu}\,
\left< \partial_t \vec{\phi}_{\bf k} \cdot \partial_t \vec{\phi}_{\bf -k}
\right> & \sim & L^{-n}\,\xi^{n+\mu-2}\,(dL/dt)^2\ ,\ \ \ \ n+\mu<2\ ,
\nonumber \\
& \sim & L^{-n}\,\ln (L/\xi)\,(dL/dt)^2\ ,\ \ \ n+\mu=2\ , \nonumber \\
& \sim & L^{\mu-2}\,(dL/dt)^2\ ,\ \ \ \ \ \ \ \ \ \ \ n+\mu>2\ .
\end{eqnarray}

In what cases does dissipation occur
primarily at scales $L(t)$, as we have just assumed? For $n>2$,
the energy density (\ref{EQN:E}), and hence dissipation,
 is dominated by variations at scale $L(t)$.
For $n \leq 2$, with $d>n$, the energy density is proportional to the defect
core volume, but the dissipation is still dominated
by scales of order $L(t)$. For this case the
rate of energy dissipation at a length scale $l$
is given by the rate at which the core volume disappears at that scale. So
the total energy dissipation is given by
\begin{equation}
\label{EQN:DISSLENGTH}
        d\epsilon/dt \sim \int dl\, n(l)\,\dot{l}\, l^{d-n-1}, \nonumber
\end{equation}
where $\dot{l}$ is the rate of evolution of defect structures of scale $l$,
$n(l)$ is the number density of features of scale $l$, and $\dot{l}\,l^{d-n-1}$
is the rate of change of the defect core volume at scale $l$.
But $j(l) \sim n(l)\,\dot{l}$ is just
the number flux of defect features, and imposing the continuity equation
 we see that $\dot{N} \sim j(0)$, where $N$ is the average
density of defects. $N$ scales
as a volume, $N \sim 1/L^d$, and so $\dot{N}$ is a constant for times of order
$L/\dot{L}$. This implies that $j(l)$ is constant for $l \ll L$, in order
to provide a constant rate of defect extinction. We see that $d\epsilon/dt
\sim \int dl \, l^{d-n-1}$ for $l \ll L$, and the convergence at small $l$
for $d>n$ implies that structures with scales $l \sim L(t)$ dominate the
energy dissipation.

For $n=d$ with $n<2$, the marginal case in the above argument,
the energy density (\ref{EQN:E}) is dominated by the core energy of
point defects and dissipation is dominated by defect pairs annihilating.
Since the dissipation occurs at separations $l \sim \xi \ll L$ we do not
expect our approach to cover these cases. In fact, since the bulk energy
density does not depend on defect
separations $l \gg \xi$, we expect the system to be disordered, with an
equilibrium density of defects, at any non-zero temperature. At $T=0$
we expect slow growth laws that depend on the details
of the potential $V(\vec{\phi})$.
These cases, $n=d$ with $n<2$, are at their lower critical dimension.

The $2d$ XY model is outside the scope of this Letter,
 since a more detailed argument \cite{Bray93c} shows that dissipation occurs
significantly at all length scales between $\xi$ and $L(t)$.
Naively, we might expect scaling violations for this case.
Yurke {\em et al.} \cite{Yurke93} suggest a growth law of $L \sim
(t/\ln{t})^{1/2}$ for  non-conserved order-parameters, while simulations by
Mondello and Goldenfeld \cite{Mondello93} find $L \sim t^{1/4}$ for the
conserved case.

In the cases where dissipation occurs at the scale of $L(t)$, we compare
the rate of energy dissipation between (\ref{EQN:RHS}) and
the time derivative of (\ref{EQN:E}) to obtain $dL/dt$
and hence $L(t)$. The results are summarized in table 1.

For non-conserved fields ($\mu=0$), we find $L
\sim t^{1/2}$ for all systems (with $d>n$ or $n>2$).
Leading corrections in the $n=2$ case are interesting: the $\ln L$
factors in (13) and (17) will in general have different effective cutoffs,
of order the core size $\xi$. This leads to a logarithmic correction
to scaling,  $L \sim t^{1/2}(1 + O(1/\ln t))$, and
may account for the smaller exponent ($\sim 0.45$) seen in
simulations of $O(2)$ systems  \cite{Toyoki91,Blundell92}.

For conserved fields ($\mu>0$) our results agree with
the the RG analysis \cite{Bray89}, with additional logarithmic
factors for the marginal cases $n=2$ and $n+\mu=2$.
We see that the conservation law is only relevant for $n+\mu \ge 2$.
Therefore for  vector fields
($n \ge 2$), any $\mu>0$ is sufficient to change the
growth law, while for scalar fields ($n=1$) the conservation law
is irrelevant for $\mu<1$, in agreement with the RG analysis \cite{Bray89}.
Simulations by Siegert and Rao
\cite{Siegert93} for $n=2$, with $d=3$ and $\mu=2$,
 obtain growth exponents slightly over
$1/4$, which may be difficult to distinguish
from  our predicted $L \sim (t\ln t)^{1/4}$ behavior.

The above discussion covers systems with purely short-ranged interactions.
We can also add long-ranged interactions to our energy functional
(\ref{EQN:HAMILTONIAN}) \cite{Hayakawa93,Bray93b}:
\begin{equation}
\label{EQN:INTERACTION}
        H_{LR} \sim \int d^dx \int d^dr
        \frac{(\vec{\phi}({\bf x}+{\bf r})-\vec{\phi}({\bf x}))^2}
                { r^{d+\sigma}}.
\end{equation}
The evaluation of the right side of
(\ref{EQN:LHSRHS}) is unchanged, but the mean energy is now given by
\begin{equation}
\label{EQN:LONGRANGE}
\epsilon_{LR} \sim \int_{\bf k} k^\sigma\,S({\bf k},t)\ ,
\ \ \ \ \ \ \sigma<2\ .
\end{equation}
We consider $\sigma < 2$ only, since for $\sigma > 2$ the energy
is dominated by short distances and we recover
the short-range $k^2$ form \cite{sigma}.
Inserting the scaling form (\ref{EQN:STRUCT}) for $S({\bf k},t)$, and
noting the asymptotic behavior (\ref{EQN:POROD}), gives
\begin{eqnarray}
\label{EQN:ELR}
\epsilon_{LR} & \sim & L^{-n}\,\xi^{n-\sigma}\ ,\ \ \ \ \ \ \ \ \ \ \ n<\sigma,
\nonumber \\
            & \sim & L^{-\sigma}\,\ln(L/\xi)\ ,\ \ \ \ \ \ \ \ \ n=\sigma,
\nonumber \\
            & \sim & L^{-\sigma}\ \ \ \ \ \ \ \ \ \ \ \ \ \ \ \ \ \  n>\sigma .
\end{eqnarray}
We then compare the rate of energy dissipation
between (\ref{EQN:RHS}), which is unchanged, and the time derivative of
(\ref{EQN:ELR}), to find $dL/dt$ and generate table 2.  Our results
agree with the leading power laws from an RG treatment by Bray \cite{Bray93b}
for the conserved case ($\mu>0$), while here we obtain the growth laws
for all cases, with any logarithmic corrections.

{}From (\ref{EQN:ELR}), we see that the long-range
interactions are relevant only for $ n \geq \sigma$  \cite{Bray93b}.
This correctly indicates that defects are non-interacting
for $n<\sigma$, where the dynamics are driven by a generalized surface
tension. Note that cases with $\sigma<0$
lead to an instability at fixed non-zero wave-vector
which breaks the scaling (\ref{EQN:STRUCT}) and hence our approach
does not directly apply. It is also interesting to note that naive power
counting using $H_{LR}$ (\ref{EQN:INTERACTION}) in the dynamics
(\ref{EQN:DYNAMICS}) yields the incorrect growth law $L \sim t^{1/\sigma}$
for non-conserved scalar fields. The fact that power
counting obtains the correct growth law, $L \sim t^{1/2}$, for short range
interactions with $n \leq 2$, can be seen as a fortuitous balancing of the UV
divergences in (\ref{EQN:GRADSQUARE}) and (\ref{EQN:TWOTIMES}).

Since systems without topological defects ($n > d+1$) will have convergent
momentum integrals for $kL \gg 1$,
we obtain $L \sim t^{1/(\sigma+\mu)}$, where $\sigma=2$ for
short-ranged forces. We can also apply this result to  systems
with topological textures ($n=d+1$), even though the appropriate Porod's
law is not known.  Since defects with $n>d$ must be spatially
extended and without a core, they will have a smaller large-$k$
tail to their structure factor $S({\bf k},t)$ than
any defects with cores.  So for $n > 2$, when the energy
dissipation clearly occurs at length scales of order $L(t)$ (see (\ref{EQN:E}))
and the momentum integrals for defects with cores converge,
our results should apply.   As a result,
tables 1 and 2 will apply for {\em any} system, except perhaps
those with $d \leq n \leq 2$ \cite{Newman90b}.
Of course, whether the structure
factor scales, as in (\ref{EQN:STRUCT}),
for systems with textures is an open question.

The strength of our approach
is that it can be applied to systems with more complicated order
parameters than $n$-component vectors.
The details of the energy functional (\ref{EQN:HAMILTONIAN})
are unimportant to this approach \cite{NOTE}.
All we need is the existence of some
short or long-ranged ``elastic energy'' ($\sigma$), an effective conservation
law ($\mu$), and the defect structure if any. Our derivation is independent
of the initial conditions, and so, e.g.,
applies equally to critical and off-critical
quenches {\em as long as the system scales at late times}.
We simply choose a Porod's law
(\ref{EQN:POROD}) to represent the dominant
defect type, which is the one represented by the smallest $n$.
For  example, in bulk  nematic liquid crystals, the existence
of string defects leads to (\ref{EQN:POROD})
with $n=2$, which with no conservation law implies a $L \sim t^{1/2}$ growth
law, consistent with recent experiments \cite{Wong92} and
simulations \cite{Blundell92}.

In summary, we obtain growth laws for phase ordering
 based only on the dynamic scaling hypothesis and a generalized Porod's
law for the large-momentum tail of the structure factor.
By focusing on the total energy dissipation, rather than the details of the
system, we do not need to consider defect dynamics. This results in a powerful
yet simple method that addresses many cases of interest.

We thank R. E. Blundell, D. A. Huse, H. H. Lee, and M. A. Moore for
discussions.

\newpage

\begin{table}
\label{TAB:SHORTRANGE}
\begin{tabular}{|l|ccc|} \hline
 $L(t)$         & $n<2$         & $n=2$         & $n>2$ \\ \hline
 $n+\mu <2$     & $t^{1/2}$     & --            & --    \\
 $n+\mu =2$     & $(t/\ln{t})^{1/2}$ & $t^{1/2}$        & --    \\
 $n+\mu >2$     & $t^{1/(n+\mu)}$ & $(t\ln{t})^{1/(2+\mu)}$& $t^{1/(2+\mu)}$ \\
\hline
\end{tabular}
\caption{The growth law of the length scale $L(t)$ for various number
of components, $n$,
and conservation laws, $\mu$.  Note that $n=d$ is excluded for $n \leq 2$.}
\end{table}

\begin{table}
\label{TAB:LONGRANGE}
\begin{tabular}{|l|ccc|} \hline
 $L(t)$         & $n< \sigma$   & $n=\sigma$    & $n>\sigma$ \\ \hline
 $n+\mu <2$     & $t^{1/2}$     & $(t \ln{t})^{1/2}$    &
$t^{1/(2+\sigma-n)}$\\
 $n+\mu =2$     & $(t/\ln{t})^{1/2}$ & $t^{1/2}$        &
                                                $(t/\ln{t})^{1/(2+\sigma-n)}$\\
 $n+\mu >2$     & $t^{1/(n+\mu)}$ & $(t\ln{t})^{1/(\sigma+\mu)}$ &
                                                        $t^{1/(\sigma+\mu)}$ \\
\hline
\end{tabular}
\caption{The growth law of the
length scale $L(t)$ for long-range forces, $0<\sigma<2$, with
various number of components, $n$,
and  conservation laws, $\mu$. Note that $n=d$ is excluded for $n \leq
\sigma$.}
\end{table}


\begin{thebibliography}{999}
\bibitem{Gunton83} For reviews see, e.g.,
J. D. Gunton, M. San Miguel and P. S. Sahni,
in {\em Phase Transitions and Critical Phenomena}, Vol.\ 8, eds.\ C. Domb
and J. L. Lebowitz (Academic, New York, 1983) p.267;
 J. S. Langer, in {\em Solids Far From
Equilibrium}, ed. C. Godr\`{e}che (Cambridge, Cambridge, 1992).

\bibitem{Mason93} N. Mason, A. N. Pargellis, and B. Yurke, Phys. Rev. Lett.
{\bf 70}, 190 (1993). These authors study a twisted nematic liquid crystal.

\bibitem{Wong92} A. P. Y. Wong, P. Wiltzius and B. Yurke, Phys.\ Rev.\ Lett.\
{\bf 68}, 3583 (1992); A. P. Y. Wong, P. Wiltzius, R. G. Larson and
B. Yurke, preprint.

\bibitem{Pargellis92} A. N. Pargellis, P. Finn, J. W. Goodby, P. Panizza,
B. Yurke, and P. E. Cladis, Phys. Rev. A {\bf 46}, 7765 (1992).

\bibitem{Toyoki91} For $n=2$:
H. Toyoki, J. Phys. Soc. Jpn. {\bf 60}, 1433 (1991);
 M. Mondello and N. Goldenfeld, Phys. Rev. A {\bf 45}, 657 (1992);
for $n=3$:
H. Toyoki, J. Phys. Soc. Jpn. {\bf 60}, 1153 (1991).

\bibitem{Blundell92}
R. E. Blundell and A. J. Bray, Phys.\ Rev.\ A {\bf 46}, R6154 (1992).
Our results indicate that nematic liquid crystals, with line
defects, should have the same growth law and leading correction as
the $O(2)$ model.

\bibitem{Mondello93} M. Mondello and N. Goldenfeld, preprint.

\bibitem{Siegert93} M. Siegert and M. Rao, preprint.

\bibitem{Yurke93} B. Yurke, A. N. Pargellis, T. Kovacs and D. A. Huse,
preprint.
See also \cite{Pargellis92}.

\bibitem{Lifshitz61} I. M. Lifshitz and V. V. Slyozov, J. Phys. Chem. Solids
{\bf 19}, 35 (1961); D. A. Huse, Phys.\ Rev.\ B {\bf 34}, 7845 (1986).

\bibitem{Lifshitz62} I. M. Lifshitz,
Zh.\ Eksp.\ Teor.\ Fiz.\ {\bf 42}, 1354 (1962)
[Sov.\ Phys.\ -- JETP {\bf 15}, 939 (1962)]; S. M. Allen and J. W. Cahn,
Acta.\ Metall.\ {\bf 27}, 1085 (1979).

\bibitem{Bray89} A. J. Bray, Phys. Rev. Lett. {\bf 62}, 2841 (1989);
        A. J. Bray, Phys.\ Rev.\ B {\bf 41}, 6724 (1990).

\bibitem{Newman90} T. J. Newman and A. J. Bray, J. Phys.\ A {\bf 23},
4491 (1990) and references therein.

\bibitem{Bray93b} A. J. Bray, preprint.

\bibitem{Liu92}F. Liu and G. F. Mazenko, Phys.\ Rev.\ B {\bf 45}, 6989 (1992);
A. J. Bray and K. Humayun, J. Phys. A {\bf 25}, 2191 (1992); H. Toyoki
and K. Honda, Prog. Theor. Phys. {\bf 78}, 237 (1987).

\bibitem{Furukawa85} H. Furukawa, Adv.\ Phys.\ {\bf 34}, 703 (1985).
See also A. Coniglio and M. Zannetti, Europhys. Lett. {\bf 10}, 575
(1989)  and A. J. Bray and K. Humayan, Phys. Rev. Lett. {\bf 68}, 1559 (1992).

\bibitem{Hohenberg77} P. C. Hohenberg and B. I. Halperin, Rev. Mod. Phys.
{\bf 49}, 435 (1977).

\bibitem{Bray93a} A. J. Bray, Phys. Rev. E {\bf 47}, 228 (1993);
A. J. Bray and K. Humayun, Phys. Rev. E {\bf 47}, R9, (1993).

\bibitem{Bray91} A. J. Bray and S. Puri, Phys. Rev. Lett. {\bf 67}, 2670
(1991).

\bibitem{Toyoki92}
H. Toyoki, Phys.\ Rev.\ B {\bf 45}, 1965 (1992).

\bibitem{Bray93c} A. J. Bray and A. D. Rutenberg, unpublished.

\bibitem{Hayakawa93} H. Hayakawa, Z. R\'{a}cz and T. Tsuzuki, preprint;
A. J. Bray, J. Phys.\ C {\bf 19}, 6225 (1986).

\bibitem{sigma} The short-range growth laws correspond to the
limit $\sigma \rightarrow 2$.

\bibitem{Newman90b} T. J. Newman, A. J. Bray, and M. A. Moore,
Phys. Rev. B {\bf 42} 4514,
 (1990) found $L \sim t^{1/4}$ for $d=1$, $n=2$, and $\mu=0$.

\bibitem{NOTE} Of course this means that our approach will not address
systems with a potential-dependent growth law, e.g. $d=n$
for $n<2$. We also do not address quenches in which thermal noise
is essential, such as systems with
static disorder (see D. A. Huse and C. L. Henley, Phys. Rev. Lett.
{\bf 54}, 2708 (1985)), or quenches to a $T>0$ critical point.

\end{thebibliography}
\end{document}